# 6G: Connectivity in the Era of Distributed Intelligence[1]

*Authors: Shilpa Talwar, Nageen Himayat, Hosein Nikopour, Feng Xue, Geng Wu, and Vida Ilderem*

**Abstract:** The confluence of 5G and AI is transforming wireless networks to deliver diverse services at the Edge, driving towards a vision of pervasive distributed intelligence. Future 6G networks will need to deliver quality of experience through seamless integration of communication, computation and AI. Therefore, networks must become intelligent, distributed, scalable, and programmable platforms across the continuum of data delivery to address the ever-increasing service requirements and deployment complexity. We present novel results across three resreach directions that are expected to be integral to 6G systems, and also discuss newer 6G metrics.

**Introduction**

The fifth generation of wireless (5G) has delivered new capabilities, targeting diverse requirements of mobile broadband, safety critical autonomous systems, and massively connected Internet of Things (IoT) applications. 5G has represented a paradigm shift from 1G to 4G, adding machine communications to co-exist along with the human centric communications. The 5G era also coincides with new breakthroughs in Artificial Intelligence (AI), which is driving sophisticated machine intelligence towards full autonomy. As we move into the 6G era, the proliferation of intelligent machines promises an age of distributed intelligence where intelligent networks and the machines they connect merge into large scale intelligent connected systems.

Significant research activity has been kicked off in academia, industry, and governments to define the vision, requirements and enabling technologies for 6G [1-5]. These include Hexa-X, a European Flagship project with bold vision of connecting human, physical, and digital worlds, RINGS a US collaborative project with focus on network resiliency through security, adaptability and/or autonomy, and ITU efforts to define network requirements for 2030.

In this paper, we pose pervasive distributed intelligence as a (sub) vision for 6G, and present how joint innovations in AI, compute and networking will be necessary to achieve it. New performance results are presented for three fundamental directions, which go beyond what is covered in previous surveys.

As we move into the 6G era, machine communications will become a significant proportion of network traffic (50% by 2023 [*Cisco Annual Internet Report*]). The data from machines will be highly diverse and will require real-time interaction and decision making. The changing nature of data (different types, sizes, dynamics) will drive changes in wireless network design. Network deployments will become increasingly dense and heterogeneous resulting in need for flexible, adaptable software-defined implementations. Managing network resources through a scalable and intelligent approach that guarantees the Quality of Service (QoS) of diverse services remains a critical challenge in 5G networks. Recent advances in Software-defined Networking (SDN) combined with Machine learning (ML) and AI tools offer potential solutions. Intelligence Defined Networking (IDN) will be the natural extension of SDN, leveraging AI/ML to add intelligence to 6G infrastructure [6].

Another key development in the 5G timeframe is transition of computing to the edge of the network to meet latency requirements of real-time applications. Edge compute will create new 'compute-centric' workload for the Radio Access Network (RAN), facilitating consumer devices with limited resources to offload their compute/memory intensive tasks to access resources at the Edge. Hence, joint consideration of Edge and RAN infrastructure, with the ability to leverage both compute/communications capabilities to meet stringent service requirements such as latency will be a key feature of 6G networks.

---

As more data is generated from IoT devices/machines, privacy concerns and data regulations can place limits on data movement. Effective data representation, processing, and local learning becomes important at every node in the data path from client to edge to cloud. A highly intelligent smart pipe will convey the semantic information or the essence of the data bits, at a rate consistent with the service requirements and the available compute-memory resources.

In this paper, we focus on technologies that promise to transform the wireless network to serve as an integrated compute-communication and learning engine. We present three cross-disciplinary research directions that lie at the intersection of traditional disciplines, and hold promise for 6G system design: (a) Intelligence Defined Networking, (b) Co-design of Compute-Communications, and (c) Distributed ML/AI, as seen Figure 1.

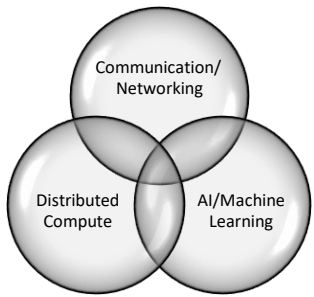

*Figure 1: 6G Cross-disciplinary Research Advancements at the intersection of disciplines.*

**Section 1: Intelligence Defined Networking (IDN)**

IDN combines AI and infrastructure programmability to enhance network operation by collecting relevant data and learning from the measured data to make predictions and decisions in an automated fashion. A study shows that IDN has the potential to increase service quality by > 50%, expand real-time services footprint by 25%, and decrease the operational costs by up to 60% [7].

Wireless network optimizations span a wide range of timescales including long term (>1s) decisions such as topology formation, mobility management of access nodes, band allocation and network slicing; medium term (>10ms) radio resource management choices like interference coordination, multi-connectivity, and traffic steering; and finally, real-time (<1ms) MAC layer decisions such as power control, user scheduling, beam tracking, etc. Intelligence in every layer of the protocol stack is required to optimize such a complex and flexible system with so many dimensions.

The advantages of ML techniques stem from the capability of learning decision-making policy for complex, dynamical systems, and the potential of creating a common framework for optimized algorithms with diverse deployments and optimization criteria. ML-based algorithms can learn complex cross-layer decision-making models, to support joint optimizations that are not easily addressed with traditional heuristic methods and have greater potential for scalability and ability to adapt to dynamic changes in the wireless system. An ML-based algorithm can, without a major redesign, be retrained to perform in specific deployments and to tune the desired optimization. Together, these benefits position ML as an enabler in emerging 6G technologies. We provide an example below to illustrate where ML-based solutions can be developed in IDN.

*Adaptative topological redundancy for resiliency and reliability*: Reliability of 6G networks is expected to improve by a factor of 10 [1]. In addition, network resiliency is the new paradigm shift in the design of the next generation wireless networks, such that the network should provide some level of service even under extreme cases where some network nodes may face attacks, failures and service disruptions [4]. Graceful performance degradation and rapid reconfiguration and recovery from disruptive events will enable 6G networks to host critical societal services, such as telemedicine, autonomous driving, etc. Furthermore, the user experience under 6G networks will be ubiquitous and with pervasive coverage across the network area. These performance requirements and capabilities demand for new technical specifications to be considered for 6G.

For example, Integrated access and backhaul (IAB) technology is already adopted in 5G, but it will need to be extended for mobile/portable nodes such as drones and even non-terrestrial communications to provide on-demand wireless infrastructure for spatially and temporally variant traffic flows. Topological redundancy is another architectural approach to improve the reliability and bring resiliency to 6G networks. Redundancy can be introduced in several forms such as UE association to multiple serving cells, multi-RAT connectivity, wireless mesh backhauls, mixed terrestrial and non-terrestrial communications, etc. Hence, 6G may introduce a reconfigurable three-dimensional multi-hop mesh network in which the topology of the network can change adaptively in two basic ways: i) location of portable access nodes, and ii) formation of the mesh network among the nodes. Notably, a multi-hop dense heterogenous network with adaptive and mobile infrastructure leads to a very dynamic interference behavior. In such a complex and adaptive network, both centralized and distributed intelligence is required to manage node mobility, network topology, cross-tier interference, and traffic steering through the multi-route multi-hop network.

Graph neural network (GNN) is a powerful ML tool to model the heterogenous wireless network and formulate such a complex RAN optimization problem. In [8], we formulate the topology formation problem as a graph optimization and propose an approach that combines deep reinforcement learning (RL) and graph embedding. The proposed approach is significantly less complex, and more scalable while it yields very close performance compared to the optimal but complex dynamic programming approach. This use case demonstrates the benefit of AI to reach the near-optimal solutions with much less complexity compared to the infeasible optimal approaches. We also use GNN to manage user-cell association for better load balancing. We consider three candidate objective functions: sum user data rate, cell coverage maximization, and load balancing. In comparison to baseline in which a user simply connects to the strongest cell, our results show up to 10% gain in throughput, 45-140% gain cell coverage, and 20- 45% gain in load balancing depending on the network deployment configurations [9].

Once the topological redundancy is enabled, advanced reliability methods such as packet-level network coding can be introduced to leverage the route diversity of the network for better resiliency. Codesign of network topology formation, network coding configuration, and traffic steering over multiple routes is a new challenge to be addressed in 6G networks.

Although ML offers a lot of potential, the application to 6G IDN networks is not without its challenges. Wireless networks are dynamic, large-scale, decentralized systems with multiple decision-making entities and suffer from real-world impairments such as measurement errors and communication delays. Large-scale datasets, which are needed to develop reliable ML-based solutions, are currently lacking in the wireless networking domain. "Out-of-the-box" ML solutions cannot simply be applied to solving real-world wireless network problems. Significant domain knowledge is needed to adapt and to further enhance existing ML solutions. In Table 1, we summarize some of the most significant challenges to incorporate ML in wireless and identify the potential for novel solution approaches for each challenge

Below we list some of the cutting-edge AI research topics that need to be developed to address the multi-tier, time-variant wireless network problems.

- *Hierarchical multi-timescale architecture*: ML agents make decisions over distinct network nodes at different layers and at different time scales leading to a hierarchical multi-timescale, multi-agent learning framework. The decisions of the agents operating at lower time scales impact the environment of the agents at the higher layer, and vice versa. Novel RL-based solutions are required to train this hierarchical structure with common objectives (see [10]).

- *Adaptive learning over graph:* In GNN a wireless network is modeled as a graph. This graph may represent a time-variant topology as network nodes move or radio connection configurations change. Adaptive and recurrent learning over a graph is required to track a time-variant wireless network. It enhances the network capability to proactively adapt itself to dynamics of the system.

- *Real-time coordinated multi-agent learning:* Performance of wireless networks improves if distributed access points coordinate in real time. The communication latency and error, however, limits the potential gain from coordination across agents. Robust multi-agent learning techniques are required which are resilient against limited communications capabilities across agents.

As we adapt ML tools for IDN design, it is essential to consider which existing tools are relevant and which tools require further development for viable ML solutions. The use of ML should not preclude the decades of work in developing model-based solutions. Hence, we must carefully consider where ML has the most potential for benefit over existing technologies.

| ML for 6G wireless IDN |||
|---|---|---|
| Cat. | Challenge | Solution space |
| Preparation of wireless for ML | Entities to collect and process wireless data not defined | Standardization for required signaling and protocols |
| | Limited public wireless network data available | Open communities to define and share reference data |
| | ML architectures not well-suited for wireless problem and data | Domain knowledge to adopt ML to wireless |
| Enhancement of ML for wireless | Several wireless optimizations in different time scales | Hierarchical AI solutions |
| | Wireless networks are dynamic and time variant | Adaptive ML to proactively optimize wireless networks |
| | Distributed ML solutions assume ideal connections between agents | Real-time multi-agent ML under wireless constraints |
| | Scalability of AI solutions for large wireless network | Combination of centralized and distributed intelligence |

*Table 1: Challenges to develop an IDN network*

**Section 2: Compute-Communication Codesign**

Traditionally compute and communication are separated as two layers in the OSI model. Applications are hosted by communications service provider (CoSP) who delivers the required service with a service level agreement. New trends have emerged over the last decade. A vast set of new applications, such as cloud gaming, distributed AI, and autonomous driving, are emerging with stringent requirements on both communication (e.g. data rate) and compute (e.g. real-time data analytics). There is also the emergence of virtualization and cloudification with generic computing platforms (e.g. Cloud-RAN, Virtual RAN, Open RAN). Given that virtualization supports both communication and compute on similar and potentially co-located platforms, there are significant codesign opportunities to address the requirements of new real-time applications. Conveniently, this aligns well with the ongoing Edge compute transition.

It is timely to support compute explicitly and natively in designing 6G, particularly at the network edge. In this new paradigm, the compute and networking schedulers will coordinate in real-time in serving user data and compute needs. Figure 2 illustrates a conceptual architecture of a co-scheduler at the Edge that jointly optimizes compute and communication resource utilization. Instead of individual schedulers slowly adapting their resources conditioned on each other, the co-scheduler will negotiate and coordinate resources based on natively shared information. As a result of co-design, a user's compute task (or a sub-task/ microservice) can be executed at an appropriate node – mobile, edge or cloud – based on the network status in real-time. At the same time, appropriate allocation of communication resources is done to support the distribution of compute task and corresponding data.

In a virtual RAN (VRAN) implementation, the communication processing, especially for complex tasks such as data decoding, is itself a part of the compute optimization. For CoSPs, the co-design with a common resource pool at edge allows resources to be scaled and adjusted as needed for serving both compute and communication needs. This leads to flexibility and efficiency similar to cloud computing.

With full state information at Edge infrastructure, codesign will allow an operator to adapt admission control, wireless scheduling and compute scheduling in real-time to meet end-to-end requirements efficiently. Here we provide a simplified model and analysis to show how more users can be supported through codesign.

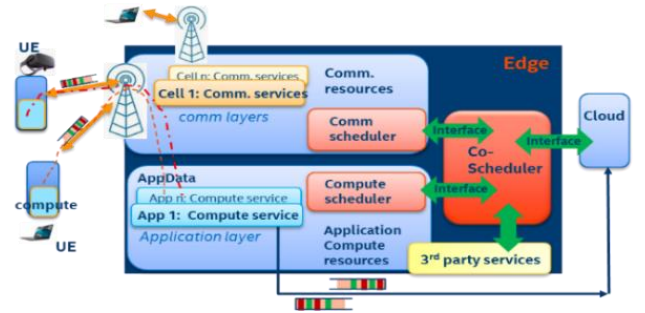

*Figure 2: High-level architecture for edge compute and communication co-design.*

***Codesign Example:*** We use a simplified example to motivate the benefits of codesign. Suppose a base-station (BS) connected to an edge server supports many mobiles with the same compute capability. Each mobile runs a video analytics application which takes its camera's input (***R0*** Mbps) and detects if there is a special event (e.g. a dog). All detection results are fed to an API on the server and the latency required is ***D*** second. Each UE (User Equipment) can split the compute into two parts, ***C1*** and ***C2***, one on the mobile with a slow CPU (Central Processing Unit) and the other on the edge with a fast CPU. For simplicity, let us assume that the detection completes whenever ***C1+C2=C***, and each user's CPU is too slow to meet the deadline by itself. Further, we assume that increasing ***C1*** helps lower the data rate from ***R0*** to ***R*** linearly such that when ***C1=C, R=0***. A traditional scheduler, at the BS would schedule its bandwidth 'fairly' considering the communication rate. Given the rate, each compute scheduler may optimize its compute split in meeting the delay $D$. Here, the users with good channels will be allocated more data rate than necessary for meeting their deadlines, while the weak users with low rate will not be able to meet their deadlines regardless of the compute split. In contrast, a co-scheduler will compensate the weak users with more wireless resource and limit the data rate of users with good channels. This simple co-design method can dramatically improve the system capacity in terms of number of admitted users meeting their QoS targets. Early results with the proposed methodology show up to 2x gains over comm-aware-only scheduling in number of users satisfying end-to-end delay constraints.

Towards enabling co-design for 6G, innovations are required in the following areas:

•*Framework for modeling and algorithm development*. It is essential to provide a unified framework for modeling comm-compute trade-offs and evaluating performance such as rate, latency, jitter, and power. The framework must encompass major application types (including AI/ML), multi-comm wireless technologies (Cellular and Wi-Fi), and heterogeneous compute platforms such as CPU, GPU, FPGA. The model also needs to consider user fairness and adaptive architectures to test variety of algorithms for resource optimization, such as a mix of AI-based and classical algorithms with partial or full telemetry,

•*Adaptive protocol, interface design and standardization*. Codesign is inherently cross-layer and operates across mobile, edge, and cloud. New interfaces are critical for a successful deployment. Several standards already exist to enhance 5G with compute support. For example, current specification of ETSI MEC [11] supports interfaces to provide close interaction between 5G core and edge computing. 6G will go significantly beyond to develop a unified framework for distributed compute and networking.

Codesign can be considered as a further evolution of network slicing with finer resolution and faster interactions between compute and communications. In ORAN, there are already interfaces specified for supporting AI-based services (e.g. *xApp*, *rApp*), for enhancing cellular operations. The co-design framework can leverage these efforts to serve both cellular and compute operations.

**Section 3: Distributed ML/AI**
Pervasive distributed intelligence mandates system capabilities to train and deploy AI solutions across all nodes in the system. To achieve vision of AI *everywhere*, we expect that 6G systems will be designed with special attention to emerging distributed AI workloads. This implies (a) distributed AI computations be aware of the constraints of the underlying wireless edge networks, and furthermore (b) wireless networks support resource optimizations specific for supporting AI computations.

Enabling Distributed AI (DAI) requires new AI solutions that continuously learn from massive data generated by intelligent machines. AI solutions that process data locally but are collaborative promise to deliver better accuracy due to accessibility of large and diverse datasets with lower privacy, bandwidth and latency cost, when compared to solutions that need to move raw data to the cloud for centralized learning. DAI when combined with continuous self-learning, and self-validating solutions holds promise to scale, fully automate, and create contextual AI solutions for virtually limitless 6G usages.

*Challenges in Enabling Distributed AI at Scale.*
DAI poses unique challenges, when compared to current centralized AI solutions. Centralized AI requires end-devices to send their collected data to a central Cloud for processing, where it can be managed jointly. In contrast, DAI requires end-devices to learn from distributed data sets locally without data sharing. The limited view of partially observed data at each device drives the need for collaboration to enhance learning accuracy. The collaboration may be assisted by a central server, as in Federated Learning, or may be decentralized and self-orchestrated by devices, as discussed in [12].

Using "Fed-Avg" algorithm, [13], as an illustrative example, Figure 3 summarizes some of the performance challenges for DAI that stem from the highly dynamic and constrained resource environment at the wireless edge. They include slow convergence, high overhead, scalability to large number of heterogeneous devices [12,13]. Additional challenges stem from an increased potential for adversarial attacks from rogue devices, privacy leakage due to sharing of model updates, and the need to self-learn from continuous data with minimal human support.

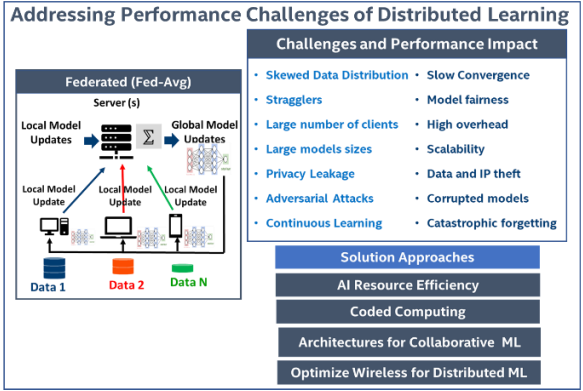

*Figure 3: Performance challenges of DAI and example solutions.*

### Addressing the Challenges in Distributed AI

DAI requires new solutions that comprehensively address the challenges of scale, performance, and automation, in combination with those of privacy and security. We highlight a few promising approaches for wireless aware distributed AI in Table 2. We also discuss how 6G systems need to be optimized to support DAI. We use benchmark datasets to illustrate our results, but the techniques are broadly applicable.

• *Optimizing Resource Efficiency for AI.* We have developed AI models that exploit awareness of compute, communication as well as data resources to address resource limitations and heterogeneity [14]. We apply statistical *importance sampling* (IS) to sample "important" devices towards speeding up the convergence of learning models and significantly reducing communication/compute costs.

Table 2 illustrates the benefits of using the importance of data in selecting devices that participate in AI training. Selecting devices based only on the speed of returning model updates can lead to significant loss in the accuracy. In contrast, "data/comm./compute" aware IS yields up to 4x improvement in training time, while improving accuracy compared to the random sampling used by Fed-Avg. The results are shown for a synthetic non-IID (independent, identically distributed) data distribution for the MNIST (Modified National Institute for Standards and Technology) hand-written character recognition dataset, as well as a naturally occurring distribution of digits across users in the Federated LEAF dataset [14]. In addition, we observed significant improvements in fairness. For example, the 10$^{th}$ percentile accuracy for the MNIST dataset is improved from 70.8% for Fed-Avg to 90.8% for IS with data and delay awareness.

| Gains from Resource-Aware Client Selection | | | | |
|---|---|---|---|---|
| | MNIST | | Federated LEAF | |
| | Mean Accuracy (%) | Training Time (Hours) | Mean Accuracy (%) | Training Time (Hours) |
| Fed-Avg | 90.7 | 12.8 (1.0x) | 76.9 | 8.2 (1.0x) |
| Data-Aware IS | 97.5 | 12.4 (1.03x) | 77.4 | 7.8 (5.1x) |
| Data-Compute-Comm-Aware IS | 96.3 | 2.8 (4.6x) | 76 | 2.1 (3.9x) |
| Gains from Coded Computing | | | | |
| | MNIST | | Fashion MNIST | |
| | Mean Accuracy (%) | Training Time (Hours) | Mean Accuracy (%) | Training Time (Hours) |
| Uncoded | 94.2 | 505 (1.0x) | 84.2 | 187 (2.7x) |
| Coded | 94.2 | 187 (2.7x) | 84.2 | 216 (2.4x) |

*Table 2: Federated Learning, Illustrative Results.*

• *Exploiting Coded Computation.* Inspired by the success of coding techniques in communication systems, we have developed a novel, coding theoretic framework to optimally add redundant computations based on compute-comm heterogeneity, to mitigate the impact of missing and straggling computations [12]. In Coded Federated Learning (CFL), each device shares a small fraction of coded data with the Server. The coded data is generated by each device privately, by taking random linear combination of the device's data. The Server can use the coded data to mitigate the effect of straggling and missing computations. This limited data sharing also combats the effects of skewed/non-IID data distributions across devices and allows for flexibility in tuning the compute workload between devices and the Server. The CFL approach provides inherent privacy for coded data, which reduces the need to add noise as a means for data obfuscation that can compromise data utility for learning. Our results for CFL with linear and non-linear regression AI models, typically yield more than 2x gains in training time compared to uncoded solutions for benchmark data sets (see Table 2), with the gains dependent on the heterogeneity of data and compute-comm. environment. Our results also show that CFL combined with IS yields up to 5x gains in total training time considering a non-IID MNIST dataset (results under submission). We note that CFL also allows for a framework to tradeoff utility, privacy and reliability, which is an area of future research.

• *New Distributed Learning Architectures.* It is also known that current FL solutions are not optimal to address non-IID data or the need to balance the computations between devices and the server. New AI architectures that leverage the capabilities of both the end-devices and the edge/cloud servers to flexibly split model training and inference between them are needed. For example, instead of all devices updating the same model, the model may be split between devices. Data and model sharing require that privacy and security concerns be balanced with performance. This leads to an inherent tradeoff in what data can be privately shared with the server and what computation may be offloaded. Striking this balance can lead to different architectural choices for DAI over wireless, and this topic is an area of active research [15].

- *Optimizing Wireless Networks for AI.* There is an opportunity to optimize wireless networks specifically for AI workloads. Exploiting wireless nodes for interim computations, as well as agents in collaborative learning, can lower the comm. and compute burden for DAI. As noted in Section 2, wireless resources can also be allocated to prioritize delay sensitive AI computations such as the gradient updates shared by the devices with the Server. Indeed, the "wireless-edge-aware" AI solutions reported in this section, can be further enhanced through joint optimization with compute and communication resource allocation approaches described in Section 2.

### Section 4: Defining New Metrics for 6G

While the typical metrics of throughput, latency & reliability will be enhanced in 6G, a broader set of key performance indicators (KPIs) will be needed in the era of distributed intelligence. Example KPI's of interest include the value & efficiency of information being communicated, network resiliency & sustainability, while scaling to satisfy "quality" for massive numbers of users. 6G also requires new integrated metrics for cross-disciplinary optimizations. For example, distributed intelligence needs metrics that measure the composite efficiency of using data, communication, and computational resources. Beyond fidelity metrics that measure the efficiency of conveying information, we must *factor the value of data/information, and* preserving "minimal sufficient statistics" when progressively processing the data in an end-to-network for an inference task. Suitable measures of *Information Efficiency* must be defined, while balancing compute, comm, and privacy cost of conveying information.

Finally, consideration of AI training may require further thought on metrics that consider complexity of AI models. Notions of *Learning Efficiency* that is indicative of the ability of AI models to generalize across new data, may also need to be considered.

**Acknowledgements:** Many fruitful discussions with NSF, academia, and US partners have helped shape our view on 6G. We also acknowledge Intel-NSF collaborations in funding US research [4,10,15].

**Author Bios:**
**Shilpa Talwar** is an Intel Fellow and director of wireless multi-communication systems with Intel Labs. She leads a research team focused on advancements in ultra-dense multi-radio network architectures and ML/AI. She is co-editor of book on 5G "Towards 5G: Applications, requirements and candidate technologies." Shilpa graduated with PhD from Stanford University, and is the author of 70 publications and 60 patents.
**Nageen Himayat** is a Director and Principal Engineer with Intel Labs, where she conducts research on distributed AI/ML and data centric protocols for 5G/6G networks. Before Intel, Nageen was with Lucent-Bell Labs and General Instrument. She obtained her Ph.D. from the University of Pennsylvania, and holds an MBA degree from University of California, Berkeley.
**Hosein Nikopour** is Manager in Intel Labs where his research focus in on intelligent wireless networks. Prior to Intel, he was with Huawei Canada where he was the inventor of Sparse Code Multiple Access (SCMA) for massive machine type communications. Hosein holds more than 60 patents and authored several publications with over 4000 citations.

**Feng Xue** is a senior research scientist with Intel Labs. His research interests include wireless communications and computing, signal processing, and information theory. He received his PhD in electrical engineering from the University of Illinois, Urbana-Champaign. He has published over 30 papers and has more than 50 patent filings.

**Geng Wu** is an Intel Fellow and Chief Technologist for Wireless Standards, leading Intel's 5G/6G development and ecosystem collaborations. He is the head of Intel's 3GPP delegation and serves as director of the board at IOWN Global Forum, MulteFire Alliance and Automotive Edge Computing Consortium.

**Vida Ilderem** is Vice President and director of Wireless Systems Research (WSR) at Intel Labs. WSR explores breakthrough wireless technologies to fulfill the promise of seamless and affordable connection and sensing for people and things. Prior to Intel, Vida served as VP of Systems and Technology Research at Motorola's Applied Research and Technology Center, where she was also recognized as Motorola Distinguished Innovator. Vida holds a doctorate and a master's degree in electrical engineering from MIT and has 27 issued patents.